\documentclass[a4paper,11pt]{article}
\usepackage{sbpo-template}
\usepackage[english]{babel}
\usepackage[latin1]{inputenc}
\usepackage{url}
\usepackage[square]{natbib}
\usepackage{indentfirst}
\usepackage{fancyhdr}
\usepackage{graphicx}
\usepackage{booktabs}
\usepackage{amsfonts}
\usepackage{amsmath, amssymb}
\usepackage{mathtools}
\usepackage[vlined, noend, ruled, linesnumbered]{algorithm2e}

\usepackage{tikz}
\usetikzlibrary{arrows}
\usetikzlibrary{arrows.meta}
\usepackage{xcolor}
\usepackage{float}
\usepackage[section]{placeins}
\usepackage[bookmarks=false]{hyperref}
\usepackage{microtype}

\newcommand{\pname}[1]{\uppercase{\texttt{#1}}}

\begin{document}

\newcommand\blfootnote[1]{%
  \begingroup
  \renewcommand\thefootnote{}\footnote{#1}%
  \addtocounter{footnote}{-1}%
  \endgroup
}

\title{Algorithms for the Maximum Eulerian Cycle Decomposition Problem\blfootnote{Anais do LIII Simp{\'o}sio Brasileiro de Pesquisa Operacional, Vol 53, 2021 - 139228}\blfootnote{\url{https://proceedings.science/sbpo-2021/}}} 

\maketitle
\thispagestyle{fancy}


\author{
\name{Pedro~Ol{\'i}mpio~Pinheiro}
\institute{Institute of Computing, University of Campinas}
\iaddress{Campinas, Brazil}
\email{pedro.pinheiro@students.ic.unicamp.br}
}

\author{
\name{Alexsandro~Oliveira~Alexandrino}
\institute{Institute of Computing, University of Campinas}
\iaddress{Campinas, Brazil}
\email{alexsandro@ic.unicamp.br}
}

\author{
\name{Andre~Rodrigues~Oliveira}
\institute{Institute of Computing, University of Campinas}
\iaddress{Campinas, Brazil}
\email{andrero@ic.unicamp.br}
}

\author{
\name{Cid~Carvalho~de~Souza}
\institute{Institute of Computing, University of Campinas}
\iaddress{Campinas, Brazil}
\email{cid@ic.unicamp.br}
}

\author{
\name{Zanoni~Dias}
\institute{Institute of Computing, University of Campinas}
\iaddress{Campinas, Brazil}
\email{zanoni@ic.unicamp.br}
}

\vspace{8mm}
\begin{resumo}
Dado um grafo Euleriano $G$, no problema de Decomposi\c c\~ao M\' axima de Ciclos em Grafos Eulerianos, estamos interessados em encontrar uma cole\c c\~ao de ciclos aresta-disjuntos $\{E_1, E_2, \ldots, E_k\}$ em $G$ tal que todas as arestas de $G$ perten\c cam a exatamente um ciclo e $k$ seja m\' aximo. Apresentamos um algoritmo para resolver o problema de \textit{pricing} da gera\c c\~ao de colunas para um modelo de Programa\c c\~ao Linear Inteira (PLI) apresentado por [Lancia e Serafini, 2016]. Al\' em disso, propomos uma heur\' istica gulosa, que busca ciclos de tamanho m\' inimo partindo de um v\' ertice aleat\' orio, e uma heur\' istica baseada na resolu\c c\~ao parcial do modelo PLI. 
Realizamos testes comparando as tr\^ es abordagens em rela\c c\~ ao \` a qualidade das solu\c c\~ oes e ao tempo de execu\c c\~ ao, usando conjuntos distintos de grafos Eulerianos aleat\' otios agrupados de acordo com a quantidade de v\' ertices e arestas.
Nossos resultados experimentais mostram que a heur\'istica baseada no modelo PLI supera os outros m\' etodos.
\end{resumo}

\bigskip
\begin{palchaves}
Decomposi\c c\~ao de Ciclos. Gera\c c\~ao de Colunas. Programa\c c\~ao Linear Inteira.

\end{palchaves}

\vspace{8mm}

\begin{abstract}
Given an Eulerian graph $G$, in the Maximum Eulerian Cycle Decomposition problem, we are interested in finding a collection of edge-disjoint cycles $\{E_1, E_2, \ldots, E_k\}$ in $G$ such that all edges of $G$ are in exactly one cycle and $k$ is maximum. We present an algorithm to solve the pricing problem of a column generation Integer Linear Programming (ILP) model introduced by \citep{lancia2016deriving}. Furthermore, we propose a greedy heuristic, which searches for minimum size cycles starting from a random vertex, and a heuristic based on partially solving the ILP model.
We performed tests comparing the three approaches in relation to the quality of solutions and execution time, using distinct sets of Eulerian graphs, each set grouping graphs with different numbers of vertices and edges.
Our experimental results show that the ILP based heuristic outperforms the other methods.
\end{abstract}

\bigskip


\section{Introduction}

The problem of Maximum Graph Decomposition consists of partitioning the set of edges of a graph into a collection {of subsets of edges} $\{E_1, E_2, \ldots, E_k\}$, such that $k$ is maximum and each subset $E_i$ induces a specific type of subgraph (e.g., trails, cycles). This is a well studied problem in combinatorial optimization~\citep{2018-ganesamurthy-paulraja, 1999-caprara, 2016-mynhardt-vanbommel, 1992-heinrich, 1990-rodger}. 

In this work, we deal with the version of the problem where each subset induces a cycle. To guarantee that a decomposition always exists, we assume that the input graph is always  Eulerian. In this way, we call this problem the \emph{Maximum Eulerian Cycle Decomposition} (\pname{MAX-ECD}). \citep{1999-caprara} showed that \pname{MAX-ECD} is NP-hard. \citep{caprara2003packing} presented an ILP model with an exponential number of variables. To cope with this, \citep{lancia2016deriving} proposed a column generation approach to solve the ILP model of \citep{caprara2003packing}. Below we discuss our motivation to study the \pname{MAX-ECD}.

In comparative genomics, the genome rearrangement distance problem receives two genomes $\mathcal{G}_1$ and $\mathcal{G}_2$ as input and consists in finding a minimum length sequence of rearrangements (mutations that alter segments of a genome) that transforms $\mathcal{G}_1$ into $\mathcal{G}_2$. One common structure used to represent genomes is the Breakpoint Graph~\citep{1999-hannenhalli-pevzner, 1995b-bafna-pevzner}. In the Breakpoint Graph, there exists one vertex for each gene of $\mathcal{G}_1$. For each pair of adjacent elements in $\mathcal{G}_1$ that are not adjacent in $\mathcal{G}_2$, there exists a black edge connecting these elements, and for each pair of adjacent elements in $\mathcal{G}_2$ that are not adjacent in $\mathcal{G}_1$, there exists a gray edge connecting these elements. 

Most algorithms for the genome rearrangement distance need the decomposition of a Breakpoint Graph into cycles of alternating colors~\citep{1995b-bafna-pevzner,1999-hannenhalli-pevzner,2008-rahman-etal}.  \citep{2020-pinheiro-etal} developed heuristics for finding decompositions of Breakpoint Graphs based on the Tabu Search metaheuristic~\citep{Glover89, Glover90}, and presented experimental results on simulated data.
\citep{1999-caprara} showed a polynomial-time reduction from the Maximum Breakpoint Graph Decomposition problem to the Maximum Eulerian Cycle Decomposition problem. Therefore, the results presented here have direct application in the genome rearrangement distance problem.

In this work, we present two heuristics for the \pname{MAX-ECD}. The first one is a greedy algorithm that, at each step, finds a minimum sized cycle starting at a random vertex and adds it to the current decomposition. The second heuristic is based on solving a version of the ILP model of \citep{caprara2003packing}, whose original size is exponential in the number of edges, by restricting it to only a few  variables. We evaluate the performance of these algorithms in practice comparing them with the column generation presented by \citep{lancia2016deriving}. 

This paper is organized as follows. Section~\ref{sec:definitions} formally introduces the problem and the definitions used in the text. Section~\ref{sec:pli} presents an integer linear programming (ILP) model for the problem. Section~\ref{sec:heuristics} describes the two heuristics we developed and Section~\ref{sec:experiments} discusses the experimental results of the ILP model and the heuristics. Finally, in Section~\ref{sec:conclusion} we highlight our conclusions and give directions for future investigations.

\section{Maximum Eulerian Cycle Decomposition}\label{sec:definitions}

A graph $G$ is an ordered pair $(V,E)$, where $V$ is a set of vertices and $E$ is a set of edges (unordered pairs of vertices). Given two vertices $u, v \in V$, we say that they are \emph{adjacent} (or \emph{neighbors}) if $(u,v) \in E$. In this work, we consider only simple graphs, i.e., graphs such that there are no duplicated edges and with $u \neq v$, for all edges $(u,v) \in E$.

We say that an edge $(u,v)$ is \emph{incident} to the vertices $u$ and $v$.
The \emph{degree} of a vertex $v$ is equal to the number of edges incident to $v$ and it is denoted by $d(v)$.

A \emph{trail} in a graph $G = (V,E)$ is a sequence of vertices $(v_1, \ldots, v_n)$, such that $(v_i, v_{i+1}) \in E$, for all $1 \leq i < n$, and there are no repetitions of edges (consecutive pairs of vertices in the trail). The \emph{length} of a trail is equal to the number of edges in it. A \emph{circuit} in a graph $G = (V,E)$ is a trail $(v_1, v_2, \ldots, v_n)$ such that $v_1 = v_n$. A \emph{cycle} is a circuit without repeated vertices. A graph is called \emph{connected} if, for all pair of vertices $u$ and $v$, there exists a trail from $u$ to $v$.

An \emph{Eulerian circuit} in a graph $G = (V,E)$ is a circuit that contains all edges of $G$. An \emph{Eulerian graph} is a graph that has an Eulerian circuit.

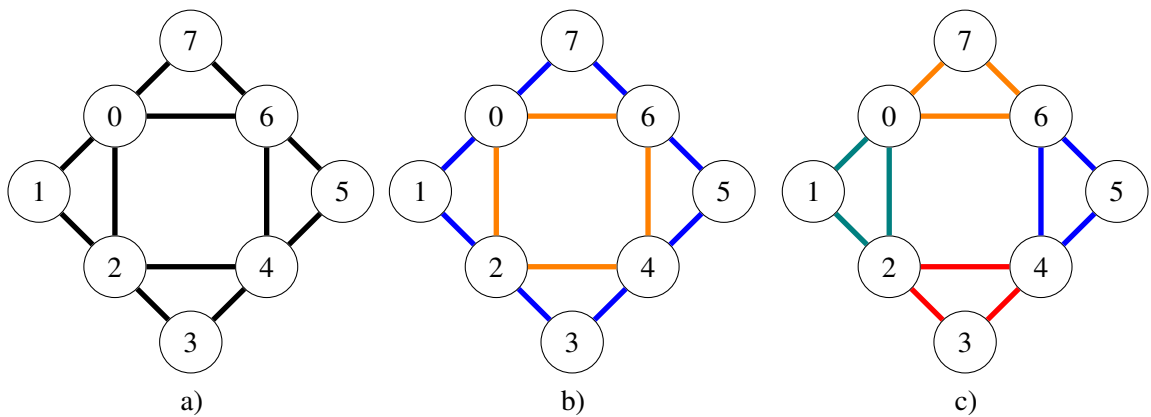
\begin{figure}[h]
\centering
\begin{minipage}{.33\textwidth}
\centering
\resizebox{1\textwidth}{!}{
\begin{tikzpicture}[scale=1]
\begin{scope}[every node/.style={inner sep=5pt, minimum size = 5pt, draw, circle}]
    \node[] (v0) at (2,1) {0};
    \node[] (v1) at (1,0) {1};
    \node[] (v2) at (2,-1) {2};
    \node[] (v3) at (3,-2) {3};
    \node[] (v4) at (4,-1) {4};
    \node[] (v5) at (5,0) {5};
    \node[] (v6) at (4,1) {6};
    \node[] (v7) at (3,2) {7};
\end{scope}

\begin{scope}[>={Stealth[black]},
              every edge/.style={draw=black, line width=2pt}]
    \path [-] (v0) edge node [black, pos=0.5, sloped, above] {} (v1);
    \path [-] (v1) edge node [black, pos=0.5, sloped, above] {} (v2);
    \path [-] (v2) edge node [black, pos=0.5, sloped, above] {} (v3);
    \path [-] (v3) edge node [black, pos=0.5, sloped, above] {} (v4);
    \path [-] (v4) edge node [black, pos=0.5, sloped, above] {} (v5);
    \path [-] (v5) edge node [black, pos=0.5, sloped, above] {} (v6);
    \path [-] (v6) edge node [black, pos=0.5, sloped, above] {} (v7);
    \path [-] (v7) edge node [black, pos=0.5, sloped, above] {} (v0);
    \path [-] (v0) edge node [black, pos=0.5, sloped, above] {} (v6);
    \path [-] (v0) edge node [black, pos=0.5, sloped, above] {} (v2);
    \path [-] (v4) edge node [black, pos=0.5, sloped, above] {} (v6);
    \path [-] (v2) edge node [black, pos=0.5, sloped, above] {} (v4);
\end{scope}
\end{tikzpicture}
}
\newline
a)
\end{minipage}%
\begin{minipage}{.33\textwidth}
\centering
\resizebox{1\textwidth}{!}{
\begin{tikzpicture}[scale=1]
\begin{scope}[every node/.style={inner sep=5pt, minimum size = 5pt, draw, circle}]
    \node[] (v0) at (2,1) {0};
    \node[] (v1) at (1,0) {1};
    \node[] (v2) at (2,-1) {2};
    \node[] (v3) at (3,-2) {3};
    \node[] (v4) at (4,-1) {4};
    \node[] (v5) at (5,0) {5};
    \node[] (v6) at (4,1) {6};
    \node[] (v7) at (3,2) {7};
\end{scope}

\begin{scope}[>={Stealth[black]},
              every edge/.style={draw=black, line width=2pt}]
    \path [-] (v0) edge [blue] node [black, pos=0.5, sloped, above] {} (v1);
    \path [-] (v1) edge [blue] node [black, pos=0.5, sloped, above] {} (v2);
    \path [-] (v2) edge [blue] node [black, pos=0.5, sloped, above] {} (v3);
    \path [-] (v3) edge [blue] node [black, pos=0.5, sloped, above] {} (v4);
    \path [-] (v4) edge [blue] node [black, pos=0.5, sloped, above] {} (v5);
    \path [-] (v5) edge [blue] node [black, pos=0.5, sloped, above] {} (v6);
    \path [-] (v6) edge [blue] node [black, pos=0.5, sloped, above] {} (v7);
    \path [-] (v7) edge [blue] node [black, pos=0.5, sloped, above] {} (v0);
    \path [-] (v0) edge [orange] node [black, pos=0.5, sloped, above] {} (v6);
    \path [-] (v0) edge [orange] node [black, pos=0.5, sloped, above] {} (v2);
    \path [-] (v4) edge [orange] node [black, pos=0.5, sloped, above] {} (v6);
    \path [-] (v2) edge [orange] node [black, pos=0.5, sloped, above] {} (v4);
\end{scope}
\end{tikzpicture}
}
\newline
b)
\end{minipage}
\hfill
\begin{minipage}{.33\textwidth}
\centering
\resizebox{1\textwidth}{!}{
\begin{tikzpicture}[scale=1]
\begin{scope}[every node/.style={inner sep=5pt, minimum size = 5pt, draw, circle}]
    \node[] (v0) at (2,1) {0};
    \node[] (v1) at (1,0) {1};
    \node[] (v2) at (2,-1) {2};
    \node[] (v3) at (3,-2) {3};
    \node[] (v4) at (4,-1) {4};
    \node[] (v5) at (5,0) {5};
    \node[] (v6) at (4,1) {6};
    \node[] (v7) at (3,2) {7};
\end{scope}

\begin{scope}[>={Stealth[black]},
              every edge/.style={draw=black, line width=2pt}]
    \path [-] (v0) edge [teal] node [black, pos=0.5, sloped, above] {} (v1);
    \path [-] (v1) edge [teal] node [black, pos=0.5, sloped, above] {} (v2);
    \path [-] (v2) edge [red] node [black, pos=0.5, sloped, above] {} (v3);
    \path [-] (v3) edge [red] node [black, pos=0.5, sloped, above] {} (v4);
    \path [-] (v4) edge [blue] node [black, pos=0.5, sloped, above] {} (v5);
    \path [-] (v5) edge [blue] node [black, pos=0.5, sloped, above] {} (v6);
    \path [-] (v6) edge [orange] node [black, pos=0.5, sloped, above] {} (v7);
    \path [-] (v7) edge [orange] node [black, pos=0.5, sloped, above] {} (v0);
    \path [-] (v0) edge [orange] node [black, pos=0.5, sloped, above] {} (v6);
    \path [-] (v0) edge [teal] node [black, pos=0.5, sloped, above] {} (v2);
    \path [-] (v4) edge [blue] node [black, pos=0.5, sloped, above] {} (v6);
    \path [-] (v2) edge [red] node [black, pos=0.5, sloped, above] {} (v4);
\end{scope}
\end{tikzpicture}
}\newline
c)
\end{minipage}
\caption{(a) Example of an Eulerian graph $G$ with $8$ vertices. (b) A cycle decomposition of $G$ into two cycles $C_1 = (0,2,4,6)$ and $C_2= (0,1,2,3,4,5,6,7)$. (c) A cycle decomposition of $G$ into four cycles $C'_1 = (0,1,2)$, $C_2 = (2,3,4)$, $C_3 = (4,5,6)$, and $C_4 = (0,6,7)$, which is a  decomposition of maximum cardinality.}
\label{fig:example}
\end{figure}

A \emph{decomposition} of a graph $G = (V, E)$ is a collection of subsets of edges $\{E_1, E_2, \ldots, E_k\}$, such that the subsets are disjoint and their union  is equal to $E$. If every subset of the collection induces a cycle in $G$, the partition forms a \emph{cycle decomposition} of $G$, as illustrated in Figure~\ref{fig:example}.

Now, given an Eulerian graph $G = (V, E)$, in the Maximum Eulerian Cycle Decomposition problem \pname{MAX-ECD}, we seek a cycle decomposition of $G$ with maximum size. Note that, since the problem is defined only for Eulerian graphs, the existence of a cycle decomposition is guaranteed.

{
Given a breadth-first tree $T = (V, E')$ of a graph $G = (V, E)$ rooted at $v \in V$, a \emph{fundamental cycle} of $T$ is any cycle that is created by inserting an edge $e \in E\setminus E'$ in $T$. 
}
This concept is used in the heuristics presented in Section~\ref{sec:heuristics}.

\section{Integer Linear Programming Model}\label{sec:pli}

Before showing the ILP model presented by \citep{caprara2003packing}, we first introduce some additional notations. We use $\mathcal{H}$ to represent the collection of cycles related to the solution constructed by the ILP model. 
Given an Eulerian graph $G = (V,E)$, let $\mathcal{C}$ be the set of all cycles in $G$. For each cycle $C \in \mathcal{C}$, let $x_C$ be a binary variable such that $x_C = 1$, if and only if $C \in \mathcal{H}$. Then, the model presented by \citep{caprara2003packing} reads:

\begin{equation}
    \label{pli-obj}
    \max \sum_{C \in \mathcal{C}} x_C
\end{equation}
subject to:
\begin{alignat}{3}
    \label{pli-constr}
    \sum_{C \in \mathcal{C} : e \in C} x_C &~\leq 1, & \forall e \in E\\
    \label{pli-integer}
    x_C & \in \{0,1\}, &~\forall C \in \mathcal{C}
\end{alignat}

Constraint~(\ref{pli-constr}) guarantees that the cycles of the solution are edge-disjoint, and equation~(\ref{pli-integer}) defines the variables $x_C$ as binary.
Note that a set of edge-disjoint cycles that do not cover all edges in the graph is a feasible solution of this model. However, such solution can not be optimal. Indeed, as the objective function~(\ref{pli-obj}) maximizes the number of cycles and $G$ is Eulerian, in any optimal solution every edge must be covered by a cycle in $\mathcal{H}$.

Since the size of $\mathcal{C}$ grows exponentially with the size of $E$, it is not viable to create all variables of the model for large instances. Therefore, we resort to column generation and branch-and-price techniques. At the beginning of the branch-and-bound  algorithm the model is fed a small subset of $x_C$ variables and more variables are added as needed until an optimal solution of the linear relaxation is reached.
The decision whether a variable is brought to the model  or not is made by computing the \emph{pricing subproblem}.
For this, consider the dual of the linear relaxation of the formulation given by equations \eqref{pli-obj}--\eqref{pli-integer}:
\begin{equation}
    \label{pli-dual-obj}
    \min \sum_{e \in E} y_e
\end{equation}
subject to:
\begin{alignat}{3}
    \label{pli-dual-constr}
    \sum_{e \in C} y_e &~\geq 1, & \forall C \in \mathcal{C}\\
    \label{pli-dual-no-negative}
    y_e~& \geq 0, &~\forall e \in E
\end{alignat}

The pricing subproblem amounts to decide if there exists a cycle $C$ whose associated constraint in~\eqref{pli-dual-constr} is violated by the current dual variables $y^*$.
Since the size of $\mathcal{C}$ is exponential in $|E|$, it is impractical to check all such constraints. 
Instead, we use a combinatorial  algorithm described below to solve the pricing subproblem.

For each cycle $C \in \mathcal{C}$, constraints~\eqref{pli-dual-constr} require that the sum of variables $y^*_e$, for $e \in C$, to be greater than or equal to $1$. Therefore, given a cycle $C$, if $\sum_{e \in C} y^*_e < 1$, the associated constraint is violated.
Hence, in the separation algorithm, we create a weighted graph $G'$ with the same vertices and edges of $G$, where each edge $e$ has weight $y^*_e$. A constraint of the dual model is violated if, and only if, the minimum weight cycle of $G'$ has weight less than one. Thus, the algorithm finds a minimum weight cycle $C^*$ in $G'$ and, if its weight is less than one, the constraint related to $C^*$ is violated and the variable $x_{C^*}$ is added to the primal model. If the weight of $C^*$ is greater than or equal to one, no variables need to be added to the linear relaxation of the primal model~\citep{lancia2016deriving}, since optimality has been reached.

So, the pricing problem is to find a minimum weight cycle $C^*$ in the weighted graph $G'$, without considering cycles related to variables $x_C$ that were fixed to zero by the branch-and-bound algorithm, since we want to find new cycles to consider in the primal model. When no variables $x_C$ are fixed to zero, we can use the \pname{MINIMUM-CYCLE} algorithm described in Algorithm~\ref{alg:minimum_cycle}. For each edge $(u,v)$, the algorithm finds a minimum weight path $P$ from $u$ to $v$, which does not have the edge $(u,v)$, and creates the cycle $P \cup {(u,v)}$. Note that this is the minimum weight cycle that contains the edge $(u,v)$. Then, the algorithm returns the generated cycle with minimum weight.

When the variables $\{x_{C_1},x_{C_2},\ldots,x_{C_k}\}$ are fixed to zero, we need to forbid these cycles in the pricing problem. For this, we create copies $G_{z_1,z_2, \ldots, z_k}$ of $G'$, such that $1 \leq z_i \leq |C_i|$ for $1 \leq i \leq k$. In each copy $G_{z_1,z_2, \ldots, z_k}$, we forbid the cycles $C_1$ to $C_k$ by updating the weight of the $z_i$-th edge of $C_i$ to $\infty$. The algorithm \pname{MINIMUM-CYCLE} is executed for each altered copy of $G'$ and the minimum weight cycle found in all these executions is the  solution of the pricing problem.

\begin{algorithm}[tb]
  \newcommand{\keyw}[1]{{\bf #1}}
  \caption{Pseudocode for finding the minimum weight cycle.\label{alg:minimum_cycle}}
  \KwIn{A graph $G'=(V,E)$ and $weight$}
  \KwOut{The cycle found ($cycle$) and its weight ($minimum$)}
  \SetAlgoLined
  $minimum \gets \infty$\\
  $cycle \gets \emptyset$\\
  \ForEach{$e = (u,v) \in E$}{
        $weight' \gets weight$\\
        $weight'[e] \gets \infty$\\
        $(w,P) \gets minimum\_weight\_path(u,v,G',weight')$\\
        $w \gets w + weight[e]$\\
        $C \gets P \cup \{e\}$\\
        \If{$w < minimum$}{
            $minimum \gets w$\\
            $cycle \gets C$\\
        }
        }
        \keyw{return} $(minimum, cycle)$
\end{algorithm}

\section{Heuristics for \pname{MAX-ECD}}\label{sec:heuristics}

In this section, we present two heuristics developed to solve the problem, which are named \pname{GREEDY} and \pname{ILP-Heuristic}.

The \pname{GREEDY} heuristic finds a cycle decomposition as described in Algorithm~\ref{alg:greedy}. At each iteration, a random vertex $v$ with degree $d(v) > 0$ is chosen. The algorithm then performs a breadth-first search to find a minimum size cycle $C$ starting at $v$, adds this cycle to the decomposition, and removes all edges of $C$ from the graph. 

Since the edges of a cycle added to the solution $\mathcal{H}$ are removed before starting a new iteration, all cycles in $\mathcal{H}$ are edge-disjoint. Note that after removing all edges of a cycle from the graph, all vertices remain with an even degree. Therefore, at each iteration, any vertex $v$ with $d(v) > 0$ is in at least one cycle. Furthermore, since the algorithm always finds a cycle, eventually $|E| = 0$, and the solution covers all edges of the original graph. The \pname{Greedy} heuristic has a time complexity of $O(|E|^2)$.

\begin{algorithm}[H]
  \newcommand{\keyw}[1]{{\bf #1}}
  \caption{Pseudocode for the {\bf \pname{Greedy}} heuristic.\label{alg:greedy}}
  \KwIn{A graph $G=(V,E)$}
  \KwOut{A cycle decomposition $\mathcal{H}$}
  \SetAlgoLined
  $\mathcal{H} = \emptyset$ \\
  \While{$|E| > 0$}{
        choose a vertex $v$ at random, such that $d(v) > 0$\\
        perform a breadth-first search to find a fundamental cycle $C$ of $v$ {of minimum size}\\
        add $C$ to $\mathcal{H}$ \\
        remove the edges of $C$ from $G$ \\
        }
        \keyw{return} $\mathcal{H}$
\end{algorithm}

The \pname{ILP-Heuristic} consists in the following. Given a set of cycles $\mathcal{C}' \subset \mathcal{C}$, we create the model from Section~\ref{sec:pli} using only variables $x_C$ such that $C \in \mathcal{C}'$. To ensure that the modified model is feasible, we can add the cycles of at least one valid decomposition in $\mathcal{C}'$. Given a parameter $k$, we create the set $\mathcal{C}'$ with the cycles of $k$ random decompositions created by the \pname{Greedy} heuristic. 

This strategy guarantees that the optimal solution of the \pname{ILP-Heuristic} is better than or equal to the best of the $k$ random solutions generated by \pname{Greedy} heuristic. 
However, a solution from the \pname{ILP-Heuristic} may not cover all edges of the input graph, since we run the ILP model using a subset of $\mathcal{C}$. Therefore, given a solution $\mathcal{H}$ from the \pname{ILP-Heuristic} for an Eulerian graph $G = (V, E)$, we apply a post-processing step with the \pname{GREEDY} heuristic that finds a decomposition $\mathcal{H}'$ for the graph $G' = (V, E')$, where $E' = \{(u,v) \in E~|~(u,v) \notin \mathcal{H}\}$, and returns $\mathcal{H} \cup \mathcal{H}'$, which is then a valid decomposition for $G$. When we refer to the \pname{ILP-Heuristic}, we assume that the post-processing step is also performed.

\section{Experimental Analysis}\label{sec:experiments}

In this section, we describe the experimental results as well as the instances generation and improvements on the execution of the ILP model.


The experiments for the \pname{MAX-ECD} were done in different sets of random Eulerian graphs, each set grouping graphs with different numbers of vertices and edges. To create these sets, we used the algorithm defined by \citep{1962:hakimi}, which is described next. Given a sequence of $n$ positive integers, the algorithm creates a graph $G$ with $n$ vertices such that there exists a one-to-one relationship between the numbers of the sequence and the degree of each vertex $v$ in $G$, or the algorithm determines that such graph does not exist. Note that the list of numbers given as a parameter to Hakimi's algorithm must have only even numbers, since we want Eulerian graphs. To generate this sequence we use Algorithm~\ref{alg:random_sequence}, that receives two parameters $m$ and $n$, such that $m \geq n$, and it returns a list $d_1, d_2, \ldots, d_n$ such that $\sum_{i=1}^n d_i = 2m$ and $d_i = 2k$, for any $1 \leq i \leq n$ and $k \in \mathbb{Z}^+$.

\begin{algorithm}[t]
  \newcommand{\keyw}[1]{{\bf #1}}
  \caption{Pseudocode for the creation of  a random sequence used by the algorithm of Hakimi.\label{alg:random_sequence}}
  \KwIn{Parameters $n$ and $m$}
  \KwOut{A list of $n$ integers $S$}
  \SetAlgoLined
  let $S$ be a list of $n$ integers of value $2$ \\
  let $m' \gets m - n$\\
  \While{$m' > 0$}{
        let $i$ be a value from $1$ to $n$ chosen uniformly at random \\
        $S_i \gets S_i + 2$\\
        $m' \gets m' - 1$\\
        }
        \keyw{return} $\mathcal{S}$
\end{algorithm}

Recall that Hakimi's algorithm may construct a disconnected graph. In this case, the following procedure is applied while the number of connected components of $G$ is greater than $1$. Choose two connected components $C$ and $C'$; then, choose at random two edges $(u,v) \in C$ and $(x,y) \in C'$; at last, replace them by the edges $(u,x)$ and $(v,y$), which decreases the number of connected components of the graph by one. Note that this procedure does not modify the degree of the vertices, and ensures that the resulting graph is connected. 

For each combination of parameters $n  \in \{10,20,\ldots, 90,100\}$ and $m \in \{0.1,\allowbreak 0.2,0.3,0.4,$ $0.5\}$, we created a set containing $20$ Eulerian graphs with $n$ vertices and $\lfloor{m}\binom{n}{2}\rfloor$  edges\footnote{Instances are available at: \href{https://github.com/compbiogroup/Maximum-Eulerian-Cycle-Decomposition-Problem}{https://github.com/compbiogroup/Maximum-Eulerian-Cycle-Decomposition-Problem}}.

\subsection{Experimental Results}

The best trade-off between time and quality of solution for the \pname{Greedy} heuristic was obtained when $k = 100$ (i.e., the algorithm was executed 100 times for each instance).
Tables~\ref{tab-sol}a-\ref{tab-sol}e show the average number of cycles returned when considering the five different densities, and Table~\ref{tab-sol}f groups these results with all densities.

In Table~\ref{tab-sol}, each entry at column \texttt{G\_avg} corresponds to the average of 100 solutions provided by the \pname{Greedy} for each instance in the corresponding set. Column \texttt{G\_max} indicates the average considering only the best decomposition returned by \pname{Greedy} for each instance.

{
We tested the ILP model with column generation described in Section~\ref{sec:pli} with different sets of initial variables and primal bounds. We tested the following sets as initial variables: all triangles (i.e., cycles with three vertices) from the graph; all fundamental cycles, which includes all triangles; and the set of cycles returned from the \pname{ILP-Heuristic}. 
}

{
We also tested the ILP model with no initial cycles, and this configuration returned the best solutions on average, whose results are in column \texttt{ILP\_cg} of Table~\ref{tab-sol}. Column \texttt{OPT} on its right indicates the proportion of instances where the algorithm has found a proven optimal solution. 
}

{
We also tested to warm start the branch-and-price algorithm with the primal bounds obtained by the heuristics \pname{GREEDY} and \pname{ILP-Heuristic}. Although, in this case,  the solver was able to find a feasible solution for all instances, 
we did not observe a relevant gain in performance. Even for instances where \texttt{ILP\_cg} had already found a solution, few or no improvements at all were observed with the strategies mentioned above. Thus, the results reported in the column \texttt{ILP\_cg} of the Table~\ref{tab-sol} correspond to the execution using no warm start.
}

Column \texttt{ILP\_h} shows the average results returned by \pname{ILP-Heuristic} described in Section~\ref{sec:pli}, using the 100 solutions returned by \pname{Greedy}, and column \texttt{OPT} on its right indicates the percentage of instances where the algorithm found an optimal solution, proved by \texttt{ILP\_cg\_h}, an ILP model with column generation using the solution found by \pname{ILP-Heuristic} as the primal bound and the cycles returned by \pname{Greedy} as initial variables.
For column \texttt{ILP\_cg}, there is a ``-'' when the ILP model was not able to finish execution and find a feasible solution for any of the instances of the set.

In most cases, the \pname{ILP-Heuristic} was able to considerably improve the $100$ random decompositions found by the \pname{Greedy} heuristic. The \pname{ILP-Heuristic} also gave better results when compared with the ILP model with column generation. Note that we use the bounds of the ILP model to prove that a solution is optimal and, for large values of $n$ and $m$, these bounds get worse, 
which could also be a reason for the low number of optimal solutions found by the \pname{ILP-Heuristic} as the value of $n$ and $m$ increases.

\begin{table}[H]
\caption{
Average number of cycles of the solutions of the following approaches: average of 100 executions of  \pname{Greedy} (\texttt{G\_avg}); maximum of 100 executions of \pname{Greedy} (\texttt{G\_max}); ILP model with column generation (\texttt{ILP\_cg}); and \pname{ILP-heuristic} (\texttt{ILP\_h}). Column \texttt{OPT} next to \texttt{ILP\_cg} has the percentage of instances that the ILP model with column generation found an optimal solution; column \texttt{OPT} next to \texttt{ILP\_h} has the percentage of instances that the \pname{ILP-Heuristic} found an optimal solution (proved by \texttt{ILP\_cg\_h}).
}

\label{tab-sol}
\begin{minipage}{0.5\linewidth}
\centering
\vspace{0.3cm}
\label{tab-sol-d10}
\resizebox{\textwidth}{!}{
\begin{tabular}{@{}|r|rr|rr|rr|@{}}
\hline
\texttt{n} & \texttt{G\_avg} & \texttt{G\_max} & \texttt{ILP\_cg} & \texttt{OPT} & \texttt{ILP\_h} & \texttt{OPT} \\ \hline
10 & 1.00 & 1.00 & 1.00 & 100\% & 1.00 & 100\% \\
20 & 1.00 & 1.00 & 1.00 & 100\% & 1.00 & 100\% \\
30 & 6.83 & 7.60 & 7.60 & 100\% & 7.60 & 100\% \\
40 & 14.46 & 16.55 & 16.85 & 95\% & 16.85 & 100\% \\
50 & 24.39 & 27.20 & 28.80 & 80\% & 28.85 & 100\% \\
60 & 37.48 & 40.95 & 43.90 & 40\% & 44.15 & 90\% \\
70 & 52.51 & 56.15 & 61.25 & 15\% & 61.85 & 40\% \\
80 & 71.19 & 75.25 & 81.95 & 0\% & 83.35 & 10\% \\
90 & 92.22 & 97.00 & - & - & 107.80 & 0\% \\
100 & 116.43 & 121.20 & - & - & 134.75 & 0\% \\ \hline
\end{tabular}
}
\vspace{0.1cm}
\newline
{\small a) Results for $m = 10\%$.}
\end{minipage}
\begin{minipage}{0.5\linewidth}
\centering
\vspace{0.3cm}
\label{tab-sol-d20}
\resizebox{\textwidth}{!}{
\begin{tabular}{@{}|r|rr|rr|rr|@{}}
\hline
\texttt{n} & \texttt{G\_avg} & \texttt{G\_max} & \texttt{ILP\_cg} & \texttt{OPT} & \texttt{ILP\_h} & \texttt{OPT} \\ \hline
10 & 1.00 & 1.00 & 1.00 & 100\% & 1.00 & 100\% \\
20 & 8.36 & 9.10 & 9.10 & 100\% & 9.10 & 100\% \\
30 & 20.90 & 22.90 & 23.85 & 95\% & 23.85 & 100\% \\
40 & 39.01 & 41.80 & 44.65 & 65\% & 44.75 & 90\% \\
50 & 63.25 & 66.60 & 71.95 & 20\% & 72.80 & 80\% \\
60 & 93.49 & 97.05 & 105.35 & 5\% & 107.20 & 60\% \\
70 & 130.26 & 134.45 & - & - & 148.40 & 0\% \\
80 & 172.83 & 177.45 & - & - & 195.50 & 0\% \\
90 & 221.70 & 227.05 & - & - & 248.50 & 0\% \\
100 & 276.52 & 282.20 & - & - & 304.25 & 0\% \\ \hline
\end{tabular}
}
\vspace{0.1cm}
\newline
{\small b) Results for $m = 20\%$.}
\end{minipage}
\begin{minipage}{0.5\linewidth}
\centering
\vspace{0.3cm}
\label{tab-sol-d30}
\resizebox{\textwidth}{!}{
\centering
\begin{tabular}{@{}|r|rr|rr|rr|@{}}
\hline
\texttt{n} & \texttt{G\_avg} & \texttt{G\_max} & \texttt{ILP\_cg} & \texttt{OPT} & \texttt{ILP\_h} & \texttt{OPT} \\ \hline
10 & 2.89 & 3.00 & 3.00 & 100\% & 3.00 & 100\% \\
20 & 14.63 & 16.00 & 16.35 & 100\% & 16.35 & 100\% \\
30 & 34.87 & 37.15 & 39.55 & 80\% & 39.65 & 100\% \\
40 & 64.74 & 67.75 & 73.65 & 50\% & 73.80 & 90\% \\
50 & 103.20 & 107.10 & 115.55 & 20\% & 117.00 & 75\% \\
60 & 151.98 & 156.35 & - & - & 172.65 & 60\% \\
70 & 209.63 & 215.05 & - & - & 237.15 & 30\% \\
80 & 277.00 & 283.35 & - & - & 308.80 & 0\% \\
90 & 354.29 & 363.20 & - & - & 394.65 & 0\% \\
100 & 441.10 & 450.90 & - & - & 487.75 & 5\% \\ \hline
\end{tabular}
}
\vspace{0.1cm}
\newline
{\small c) Results for $m = 30\%$.}
\end{minipage}
\begin{minipage}{0.5\linewidth}
\centering
\vspace{0.3cm}
\label{tab-sol-d40}
\resizebox{\textwidth}{!}{
\centering
\begin{tabular}{@{}|r|rr|rr|rr|@{}}
\hline
\texttt{n} & \texttt{G\_avg} & \texttt{G\_max} & \texttt{ILP\_cg} & \texttt{OPT} & \texttt{ILP\_h} & \texttt{OPT} \\ \hline
10 & 4.83 & 5.05 & 5.05 & 100\% & 5.05 & 100\% \\
20 & 21.04 & 22.70 & 23.65 & 85\% & 23.70 & 100\% \\
30 & 49.51 & 52.00 & 55.95 & 80\% & 56.10 & 100\% \\
40 & 90.54 & 93.75 & 101.30 & 50\% & 102.05 & 100\% \\
50 & 144.30 & 148.65 & - & - & 162.35 & 100\% \\
60 & 210.99 & 216.65 & - & - & 235.45 & 100\% \\
70 & 290.41 & 297.95 & - & - & 321.75 & 100\% \\
80 & 382.82 & 393.40 & - & - & 420.95 & 95\% \\
90 & 487.99 & 500.85 & - & - & 533.25 & 55\% \\
100 & 606.16 & 622.10 & - & - & 658.35 & 10\% \\ \hline
\end{tabular}
}
\vspace{0.1cm}
\newline
{\small d) Results for $m = 40\%$.}
\end{minipage}
\begin{minipage}{0.5\linewidth}
\centering
\vspace{0.3cm}
\label{tab-sol-d50}
\resizebox{\textwidth}{!}{
\centering
\begin{tabular}{@{}|r|rr|rr|rr|@{}}
\hline
\texttt{n} & \texttt{G\_avg} & \texttt{G\_max} & \texttt{ILP\_cg} & \texttt{OPT} & \texttt{ILP\_h} & \texttt{OPT} \\ \hline
10 & 6.05 & 6.40 & 6.40 & 100\% & 6.40 & 100\% \\
20 & 27.39 & 29.20 & 30.75 & 100\% & 30.75 & 100\% \\
30 & 63.88 & 66.50 & 71.20 & 60\% & 71.75 & 100\% \\
40 & 116.56 & 120.30 & 128.55 & 35\% & 129.70 & 100\% \\
50 & 185.03 & 190.50 & - & - & 203.95 & 100\% \\
60 & 269.97 & 277.30 & - & - & 295.00 & 100\% \\
70 & 370.75 & 380.45 & - & - & 402.00 & 100\% \\
80 & 488.34 & 500.95 & - & - & 526.00 & 100\% \\
90 & 621.70 & 637.95 & - & - & 666.75 & 75\% \\
100 & 771.67 & 791.85 & - & - & 823.25 & 30\% \\ \hline
\end{tabular}
}
\vspace{0.1cm}
\newline
{\small e) Results for $m = 50\%$.}
\end{minipage}
\begin{minipage}{0.5\linewidth}
\centering
\vspace{0.3cm}
\label{tab-sol-all}
\resizebox{\textwidth}{!}{
\centering
\begin{tabular}{@{}|r|rr|rr|rr|@{}}
\hline
\texttt{n} & \texttt{G\_avg} & \texttt{G\_max} & \texttt{ILP\_cg} & \texttt{OPT} & \texttt{ILP\_h} & \texttt{OPT} \\ \hline
10 & 3.15 & 3.29 & 3.29 & 100\% & 3.29 & 100\% \\
20 & 14.48 & 15.60 & 16.17 & 97\% & 16.18 & 100\% \\
30 & 35.20 & 37.23 & 39.63 & 83\% & 39.79 & 100\% \\
40 & 65.06 & 68.03 & 73.00 & 59\% & 73.43 & 95\% \\
50 & 104.03 & 108.01 & - & - & 116.99 & 91\% \\
60 & 152.78 & 157.66 & - & - & 170.89 & 82\% \\
70 & 210.71 & 216.81 & - & - & 234.23 & 53\% \\
80 & 278.44 & 286.08 & - & - & 306.92 & 41\% \\
90 & 355.58 & 365.21 & - & - & 390.19 & 26\% \\
100 & 442.37 & 453.65 & - & - & 481.67 & 9\% \\ \hline
\end{tabular}
}
\vspace{0.1cm}
\newline
{\small f) Results for all values of $m$.}
\end{minipage}
\end{table}

\begin{figure}[t]
    \centering
    \begin{minipage}{\linewidth}
    \centering 
     \includegraphics[width=0.6\textwidth]{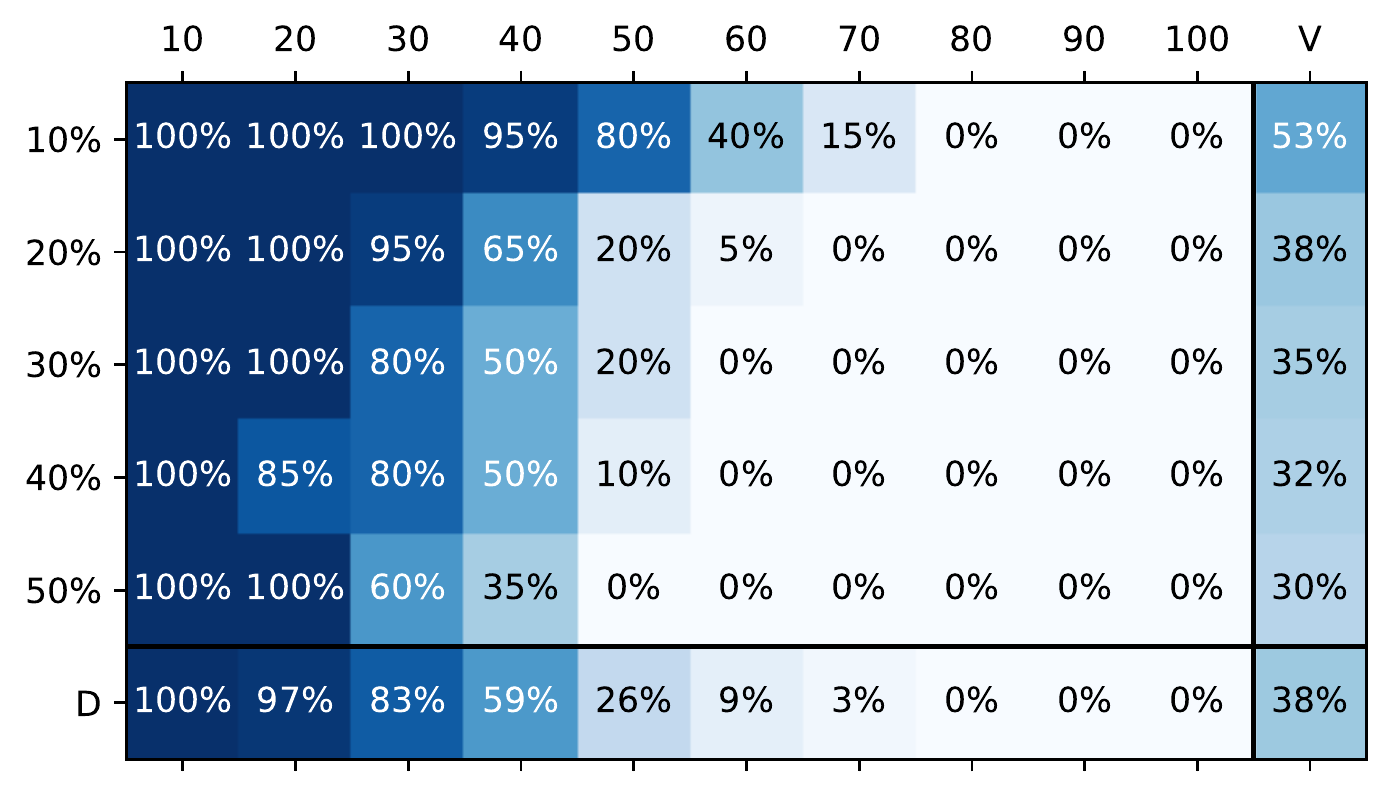}\label{image_optimal_solutions_pli}
    \end{minipage}
    \newline
     {a) ILP model with column generation (\texttt{ILP\_cg})}
    \begin{minipage}{\linewidth}
    \centering 
     \includegraphics[width=0.6\textwidth]{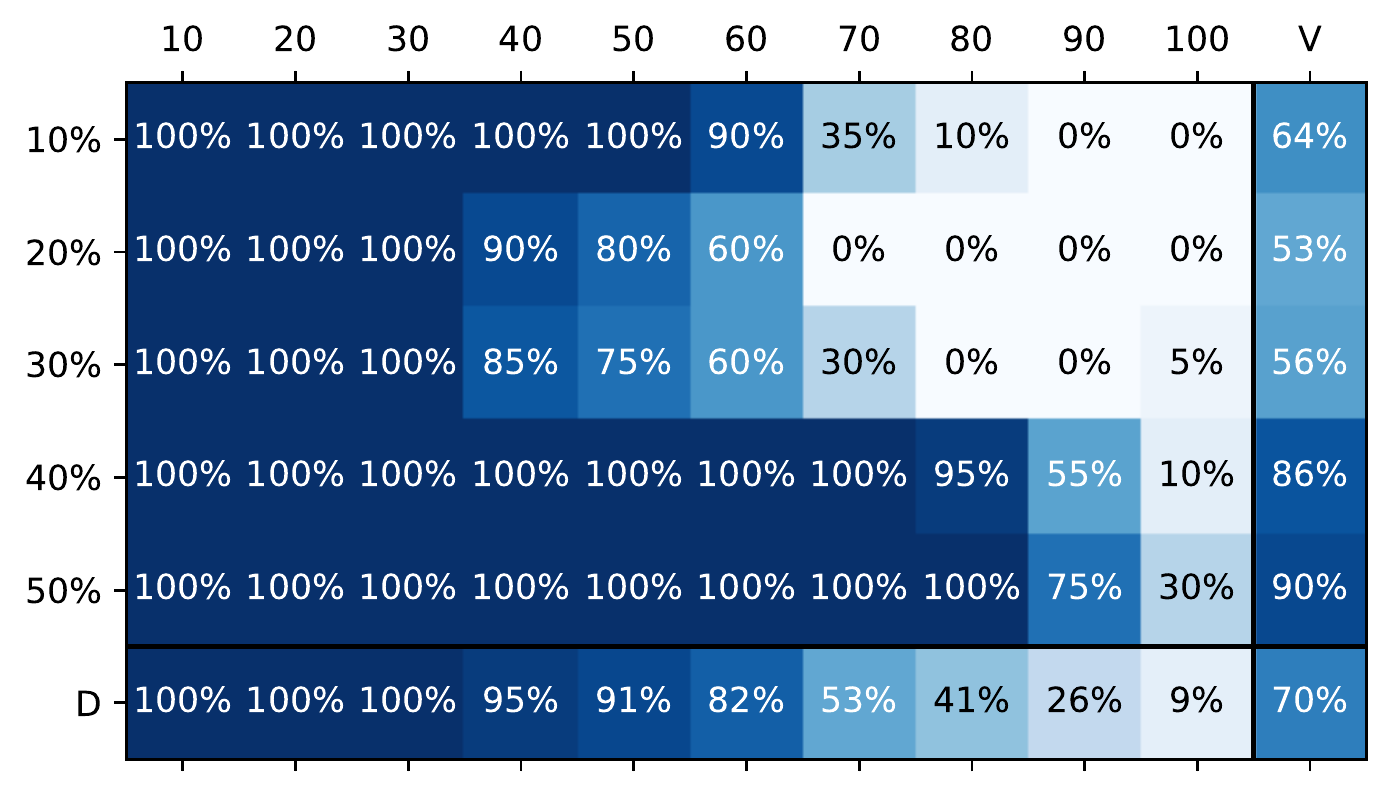}\label{image_optimal_solutions_pli_h}
    \end{minipage} 
    \newline
     {b) \pname{ILP-Heuristic} (\texttt{ILP\_h})}
     \caption{{The horizontal axis indicates the number of vertices of the instances, and the vertical axis indicates the density of the graph. Each cell indicates {\bf (a)} the percentage of optimal solutions found by the ILP model with column generation and {\bf (b)} the percentage of optimal solutions found by the \pname{ILP-Heuristic} for each set of instances. Line D and column V show the percentage of optimal solutions grouped by density and by number of vertices, respectively.}\label{image_optimal_solutions}}
\end{figure}

{Figure~\ref{image_optimal_solutions} shows the percentage of optimal solutions found by the ILP model with column generation (\texttt{ILP\_cg}) and by the \pname{ILP-Heuristic}. We can see that the \pname{ILP-Heuristic} was significantly better on finding optimal solutions in this experiment, finding optimal solutions in 69.7\% of the instances, while \texttt{ILP\_cg} found the optimal solution only on 37.7\% of them.
We can observe that the number of optimal solutions from \pname{ILP-Heuristic} increases for graphs with a higher number of edges.} This can be explained by the fact that the denser the graphs, the greater the chances of having multiple cycles of size 3 (called triangles). Table~\ref{table:triangles} shows the percentage of triangles in the solutions returned by the \pname{ILP-Heuristic}, where we can observe a clear trend of this type of cycle as the density increases. For instance, while the percentage of triangles is $29.96\%$ when $m=10\%$, this percentage grows to $95.40\%$ when $m=50\%$.

\begin{table}[t]
\caption{Percentage of the number of cycles of size $3$ (triangles) in the solutions returned by the \pname{ILP-Heuristic}.\label{table:triangles}}
\centering
\vspace{0.3cm}
{\def\arraystretch{1.0}\tabcolsep=6pt
\begin{tabular}{|r|rrrrr|}
\hline
\texttt{n} & $m=10$\% & $m=20$\% & $m=30$\% & $m=40$\% & $m=50$\% \\ \hline
10 & 0.00\% & 0.00\% & 31.67\% & 59.41\% & 67.19\% \\
20 & 0.00\% & 36.81\% & 63.30\% & 81.43\% & 91.87\% \\
30 & 24.34\% & 53.46\% & 75.13\% & 89.19\% & 97.63\% \\
40 & 27.89\% & 60.65\% & 83.40\% & 94.27\% & 99.31\% \\
50 & 34.09\% & 67.29\% & 86.40\% & 98.18\% & 99.93\% \\
60 & 37.41\% & 72.42\% & 92.56\% & 99.41\% & 100.00\% \\
70 & 38.54\% & 75.57\% & 94.89\% & 99.77\% & 99.75\% \\
80 & 42.74\% & 77.85\% & 95.34\% & 99.30\% & 99.33\% \\
90 & 46.10\% & 81.34\% & 97.37\% & 99.45\% & 99.45\% \\
100 & 48.53\% & 83.44\% & 98.45\% & 99.45\% & 99.50\% \\ \hline
\end{tabular}
}
\end{table}

Tables~\ref{tab-time}a-\ref{tab-time}e show the average time (in seconds) spent by the algorithms when considering the five different densities, and Table~\ref{tab-time}f groups these results with all densities. The time limit was set to 1800 seconds for every algorithm.

In the cases \texttt{ILP\_cg\_h} was able to prove the optimality, this was done in 20 seconds on average, and in only 1.9\% of cases it needed more than 60 seconds. In the remaining cases, \texttt{ILP\_cg\_h} was not able to prove optimality nor to improve the solution obtained by \pname{ILP-Heuristic}, within the limit of 1800 seconds.

Among the cases where it was not possible to prove optimality for the solution obtained by the \pname{ILP-Heuristic}, in 68.0\% of them neither \pname{ILP-Heuristic} nor \texttt{ILP\_cg\_h} managed to obtain an optimal solution within the limit of 1800 seconds. In the other 32.0\%, \pname{ILP-Heuristic} found an optimal solution (considering the decompositions given as input), but \texttt{ILP\_cg\_h} was unable to prove optimality within the time limit.

All the experiments were conducted on a PC equipped with a 2.4GHz Intel 
Core\texttrademark~i7-8700T, with 12 cores and 8 GB of RAM, running Ubuntu 18.04.3. We used SCIP 6.0.2~\citep{achterberg2009scip} as the ILP solver and Gurobi 8.1.1~\citep{gurobi} as the LP solver. Since the \pname{ILP-Heuristic} does not need column generation, we used Gurobi as the ILP solver for this heuristic.

\section{Conclusion}\label{sec:conclusion}

In this work we presented algorithms for the Maximum Eulerian Cycle Decomposition problem, alongside an experimental study of these algorithms and a column generation ILP model for the problem. 

Our experiments showed good results for the ILP heuristic, which is an heuristic that partially solves the ILP model activating only the variables related to good cycles found by a greedy heuristic.

For future work, we intend to propose new algorithms based on metaheuristics used in the literature for combinatorial problems, such as Genetic Algorithms (GA) and 
Greedy Randomized Adaptive Search Procedure (GRASP). 

\begin{table}[t]
\caption{Average running time in seconds of the following approaches: 100 executions of \pname{GREEDY} (\texttt{G\_avg}); \pname{ILP-Heuristic} (\texttt{ILP\_h}); ILP model with column generation (\texttt{ILP\_cg}); and the ILP model with column generation using the solution found by \pname{ILP-Heuristic} and the cycles returned by \pname{GREEDY} (\texttt{ILP\_cg\_h}).
}
\label{tab-time}
\begin{minipage}{0.5\linewidth}
\centering
\vspace{0.3cm}
\label{tab-time-d10}
\resizebox{\textwidth}{!}{
\centering
\begin{tabular}{@{}|r|rrrr|@{}}
\hline
\texttt{n} & \texttt{G\_avg} & \texttt{ILP\_h} & \texttt{ILP\_cg} & \texttt{ILP\_cg\_h} \\ \hline
10 & 0.00 & 0.00 & 0.00 & 0.00 \\
20 & 0.01 & 0.00 & 0.00 & 0.00 \\
30 & 0.21 & 0.00 & 0.00 & 0.00 \\
40 & 1.02 & 0.02 & 90.05 & 0.06 \\
50 & 3.89 & 0.23 & 391.90 & 0.20 \\
60 & 6.97 & 1.55 & 1082.40 & 218.03 \\
70 & 11.42 & 6.20 & 1629.85 & 1189.97 \\
80 & 17.28 & 82.59 & 1800.00 & 1620.33 \\
90 & 35.56 & 752.49 & 1800.00 & 1800.00 \\
100 & 48.12 & 1800.00 & 1800.00 & 1800.00 \\ \hline
\end{tabular}
}
\vspace{0.1cm}
\newline
{\small a) Results for $m = 10\%$.}
\end{minipage}
\begin{minipage}{0.5\linewidth}
\centering
\vspace{0.3cm}
\label{tab-time-d20}
\resizebox{\textwidth}{!}{
\centering
\begin{tabular}{@{}|r|rrrr|@{}}
\hline
\texttt{n} & \texttt{G\_avg} & \texttt{ILP\_h} & \texttt{ILP\_cg} & \texttt{ILP\_cg\_h} \\ \hline
10 & 0.00 & 0.00 & 0.00 & 0.00 \\
20 & 0.08 & 0.00 & 0.00 & 0.00 \\
30 & 0.55 & 0.06 & 116.90 & 0.03 \\
40 & 1.94 & 0.63 & 635.40 & 230.08 \\
50 & 5.31 & 4.57 & 1525.60 & 369.83 \\
60 & 12.09 & 113.80 & 1786.70 & 722.72 \\
70 & 23.06 & 826.49 & 1800.00 & 1800.00 \\
80 & 41.02 & 1709.79 & 1800.00 & 1800.00 \\
90 & 68.08 & 1800.00 & 1800.00 & 1800.00 \\
100 & 107.77 & 1800.00 & 1800.00 & 1800.00 \\ \hline
\end{tabular}
}
\vspace{0.1cm}
\newline
{\small b) Results for $m = 20\%$.}
\end{minipage}
\begin{minipage}{0.5\linewidth}
\centering
\vspace{0.3cm}
\label{tab-time-d30}
\resizebox{\textwidth}{!}{
\centering
\begin{tabular}{@{}|r|rrrr|@{}}
\hline
\texttt{n} & \texttt{G\_avg} & \texttt{ILP\_h} & \texttt{ILP\_cg} & \texttt{ILP\_cg\_h} \\ \hline
10 & 0.01 & 0.00 & 0.00 & 0.00 \\
20 & 0.15 & 0.01 & 0.00 & 0.01 \\
30 & 0.83 & 0.22 & 373.80 & 62.96 \\
40 & 3.13 & 2.24 & 1071.05 & 360.14 \\
50 & 9.21 & 24.01 & 1558.15 & 494.85 \\
60 & 20.90 & 594.01 & 1800.00 & 906.69 \\
70 & 42.67 & 1432.95 & 1800.00 & 1261.47 \\
80 & 79.32 & 1800.00 & 1800.00 & 1800.00 \\
90 & 139.11 & 1800.00 & 1800.00 & 1800.00 \\
100 & 230.98 & 1749.13 & 1800.00 & 1710.90 \\ \hline
\end{tabular}
}
\vspace{0.1cm}
\newline
{\small c) Results for $m = 30\%$.}
\end{minipage}
\begin{minipage}{0.5\linewidth}
\centering
\vspace{0.3cm}
\label{tab-time-d40}
\resizebox{\textwidth}{!}{
\centering
\begin{tabular}{@{}|r|rrrr|@{}}
\hline
\texttt{n} & \texttt{G\_avg} & \texttt{ILP\_h} & \texttt{ILP\_cg} & \texttt{ILP\_cg\_h} \\ \hline
10 & 0.01 & 0.00 & 0.00 & 0.00 \\
20 & 0.21 & 0.02 & 270.00 & 3.84 \\
30 & 1.31 & 0.69 & 415.10 & 28.32 \\
40 & 5.31 & 5.41 & 1054.95 & 0.29 \\
50 & 15.63 & 19.72 & 1697.00 & 0.98 \\
60 & 37.36 & 146.00 & 1800.00 & 20.71 \\
70 & 77.82 & 107.12 & 1800.00 & 4.51 \\
80 & 150.90 & 620.56 & 1800.00 & 99.14 \\
90 & 268.44 & 998.60 & 1800.00 & 822.75 \\
100 & 454.00 & 1686.91 & 1800.00 & 1623.69 \\ \hline
\end{tabular}
}
\vspace{0.1cm}
\newline
{\small d) Results for $m = 40\%$.}
\end{minipage}
\begin{minipage}{0.5\linewidth}
\centering
\vspace{0.3cm}
\label{tab-time-d50}
\resizebox{\textwidth}{!}{
\centering
\begin{tabular}{@{}|r|rrrr|@{}}
\hline
\texttt{n} & \texttt{G\_avg} & \texttt{ILP\_h} & \texttt{ILP\_cg} & \texttt{ILP\_cg\_h} \\ \hline
10 & 0.01 & 0.00 & 0.00 & 0.00 \\
20 & 0.30 & 0.03 & 4.75 & 0.02 \\
30 & 2.05 & 0.29 & 742.80 & 0.10 \\
40 & 8.61 & 3.37 & 1243.40 & 0.37 \\
50 & 25.35 & 17.44 & 1800.00 & 1.14 \\
60 & 62.28 & 69.24 & 1800.00 & 2.65 \\
70 & 134.90 & 325.33 & 1800.00 & 6.01 \\
80 & 265.77 & 1266.17 & 1800.00 & 14.29 \\
90 & 487.29 & 1428.34 & 1800.00 & 474.75 \\
100 & 902.33 & 1665.81 & 1800.00 & 1273.16 \\ \hline
\end{tabular}
}
\vspace{0.1cm}
\newline
{\small e) Results for $m = 50\%$.}
\end{minipage}
\begin{minipage}{0.5\linewidth}
\centering
\vspace{0.3cm}
\label{tab-time-all}
\resizebox{\textwidth}{!}{
\centering
\begin{tabular}{@{}|r|rrrr|@{}}
\hline
\texttt{n} & \texttt{G\_avg} & \texttt{ILP\_h} & \texttt{ILP\_cg} & \texttt{ILP\_cg\_h} \\ \hline
10 & 0.01 & 0.00 & 0.00 & 0.00 \\
20 & 0.15 & 0.02 & 54.95 & 0.77 \\
30 & 0.99 & 0.25 & 329.72 & 18.28 \\
40 & 4.00 & 2.33 & 818.97 & 118.19 \\
50 & 11.88 & 13.20 & 1394.53 & 173.40 \\
60 & 27.92 & 184.92 & 1653.82 & 374.16 \\
70 & 57.97 & 539.62 & 1765.97 & 852.39 \\
80 & 110.86 & 1095.82 & 1800.00 & 1066.75 \\
90 & 199.70 & 1355.89 & 1800.00 & 1339.50 \\
100 & \phantom{0}348.64 & 1740.37 & 1800.00 & 1641.55 \\ \hline
\end{tabular}
}
\vspace{0.1cm}
\newline
{\small f) Results for all values of $m$.}
\end{minipage}

\end{table}

\paragraph{Acknowledgments.}
This work was supported by the National Council of Technological and Scientific Development, CNPq (425340/2016-3 and 304380/2018-0),
the Coordena\c c\~ ao de Aperfei\c coamento de Pessoal de N\' ivel Superior - Brasil (CAPES) - Finance Code 001
, and the S\~ao Paulo Research Foundation, FAPESP (grants %
2013/08293-7
, 2015/11937-9
, 2017/12646-3
, 2019/25410-3
, and 2019/27331-3
).
 

\begin{thebibliography}{16}
\expandafter\ifx\csname natexlab\endcsname\relax\def\natexlab#1{#1}\fi
\expandafter\ifx\csname url\endcsname\relax
  \def\url#1{{\tt #1}}\fi

\bibitem[Achterberg(2009)]{achterberg2009scip}
Achterberg, T. (2009).
\newblock {SCIP: Solving Constraint Integer Programs}.
\newblock {\em Mathematical Programming Computation}, 1\penalty0 (1):\penalty0
  1--41.

\bibitem[Bafna and Pevzner(1995)]{1995b-bafna-pevzner}
Bafna, V. and Pevzner, P.~A. (1995).
\newblock {Sorting Permutations by Transpositions}.
\newblock In {\em {Proceedings of the 6th Annual ACM-SIAM Symposium on Discrete
  Algorithms (SODA'1995)}}, p. 614--623, Philadelphia, PA, USA. Society for
  Industrial and Applied Mathematics.

\bibitem[Caprara(1999)]{1999-caprara}
Caprara, A. (1999).
\newblock {Sorting Permutations by Reversals and Eulerian Cycle
  Decompositions}.
\newblock {\em {SIAM Journal on Discrete Mathematics}}, 12\penalty0
  (1):\penalty0 91--110.

\bibitem[Caprara et~al.(2003)Caprara, Panconesi, and Rizzi]{caprara2003packing}
Caprara, A., Panconesi, A., and Rizzi, R. (2003).
\newblock {Packing Cycles in Undirected Graphs}.
\newblock {\em Journal of Algorithms}, 48\penalty0 (1):\penalty0 239--256.

\bibitem[Ganesamurthy and Paulraja(2018)]{2018-ganesamurthy-paulraja}
Ganesamurthy, S. and Paulraja, P. (2018).
\newblock {2p-Cycle Decompositions of Some Regular Graphs and Digraphs}.
\newblock {\em Discrete Mathematics}, 341\penalty0 (8):\penalty0 2197--2210.

\bibitem[Glover(1989)]{Glover89}
Glover, F.~W. (1989).
\newblock Tabu search - part {I}.
\newblock {\em {INFORMS} Journal on Computing}, 1\penalty0 (3):\penalty0
  190--206.

\bibitem[Glover(1990)]{Glover90}
Glover, F.~W. (1990).
\newblock Tabu search - part {II}.
\newblock {\em {INFORMS} Journal on Computing}, 2\penalty0 (1):\penalty0 4--32.

\bibitem[{Gurobi Optimization, LLC}(2020)]{gurobi}
{Gurobi Optimization, LLC} (2020).
\newblock {Gurobi Optimizer Reference Manual}.
\newblock \url{https://www.gurobi.com}.
\newblock Access in November 23, 2020.

\bibitem[Hakimi(1962)]{1962:hakimi}
Hakimi, S.~L. (1962).
\newblock {On Realizability of a Set of Integers as Degrees of the Vertices of
  a Linear Graph. I}.
\newblock {\em Journal of the Society for Industrial and Applied Mathematics},
  10\penalty0 (3):\penalty0 496--506.

\bibitem[Hannenhalli and Pevzner(1999)]{1999-hannenhalli-pevzner}
Hannenhalli, S. and Pevzner, P.~A. (1999).
\newblock {Transforming Cabbage into Turnip: Polynomial Algorithm for Sorting
  Signed Permutations by Reversals}.
\newblock {\em {Journal of the ACM}}, 46\penalty0 (1):\penalty0 1--27.

\bibitem[Heinrich(1992)]{1992-heinrich}
Heinrich, K. (1992).
\newblock {Path Decomposition}.
\newblock {\em Le Matematiche}, 47\penalty0 (2):\penalty0 241--258.

\bibitem[Lancia and Serafini(2016)]{lancia2016deriving}
Lancia, G. and Serafini, P. (2016).
\newblock {Deriving Compact Extended Formulations via LP-based Separation
  Techniques}.
\newblock {\em Annals of Operations Research}, 240\penalty0 (1):\penalty0
  321--350.

\bibitem[Mynhardt and van Bommel(2016)]{2016-mynhardt-vanbommel}
Mynhardt, C.~M. and van Bommel, C.~M. (2016).
\newblock {Triangle Decompositions of Planar Graphs}.
\newblock {\em Discussiones Mathematicae Graph Theory}, 36\penalty0
  (3):\penalty0 643--659.

\bibitem[Pinheiro et~al.(2020)Pinheiro, Alexandrino, Oliveira, de~Souza, and
  Dias]{2020-pinheiro-etal}
Pinheiro, P.~O., Alexandrino, A.~O., Oliveira, A.~R., de~Souza, C.~C., and
  Dias, Z. (2020).
\newblock {Heuristics for Breakpoint Graph Decomposition with Applications in
  Genome Rearrangement Problems}.
\newblock In {\em {Proceedings of the 13th Brazilian Symposium on
  Bioinformatics (BSB'2020)}}, p. 129--140. {Springer International
  Publishing}.

\bibitem[Rahman et~al.(2008)Rahman, Shatabda, and Hasan]{2008-rahman-etal}
Rahman, A., Shatabda, S., and Hasan, M. (2008).
\newblock {An Approximation Algorithm for Sorting by Reversals and
  Transpositions}.
\newblock {\em {Journal of Discrete Algorithms}}, 6\penalty0 (3):\penalty0
  449--457.

\bibitem[Rodger(1990)]{1990-rodger}
Rodger, C.~A. (1990).
\newblock {Graph Decomposition}.
\newblock {\em Le Matematiche}, 45\penalty0 (1):\penalty0 119--140.

\end{thebibliography}

\end{document}